\documentclass{pasj00}
\draft

\usepackage{threeparttable}
\begin{document}
\SetRunningHead{Author(s) in page-head}{Running Head}
\Received{2010/02/11}
\Accepted{; version 0.0 May. 9, 2010}

\title{A Search for Water Masers in the Saturnian System}

\author{Shigeru T\textsc{akahashi} %
}
\email{shigeru@nro.nao.ac.jp}

\author{Shuji D\textsc{eguchi},  Nario K\textsc{uno}, Tomomi S\textsc{himoikura}\thanks{present address: Department of Astronomy and Earth Sciences, Tokyo Gakugei University, 4-1-1 Nukui-Kita-Machi, Koganei, Tokyo 184-8501, Japan}}


\affil{Nobeyama Radio Observatory, National Astronomical Observatory of Japan\\
462-2 Nobeyama, Minamimaki, Minamisaku, Nagano 384-1305, Japan}


\and
\author{Fumi Y\textsc{oshida}}
\affil{National Astronomical Observatory of Japan\\
2-21-1, Osawa, Mitaka, Tokyo 181-8588, Japan}

%


\maketitle

\KeyWords{planets and satellites: individual Titan, Hyperion, Iapetus, Enceladus and Atlas} 

\begin{abstract}
We searched for ${\rm H_{2}O}$ 6(1,6)-5(2,3) maser emission at 22.235 GHz from several Saturnian satellites with the Nobeyama 45m radio telescope  in May 2009.  Observations were made for Titan, Hyperion, Enceladus and Atlas, for which Pogrebenko {\it et al.} (2009) had reported detections of water masers at 22.235 GHz, and in addition for Iapetus and  other inner satellites. We detected  no emission of the water maser line for all the satellites observed, 
although sensitivities of our observations were comparable or even better than those of  Pogrebenko {\it et al.}. 
We infer that the water maser emission from the Saturnian system is extremely weak, or  sporadic in nature.  Monitoring over a long period and obtaining statistical results must be made for the further understanding of the water maser emission in the Saturnian system.
\end{abstract}

\section{Introduction}

Maser emissions are widely found in celestial objects such as dense cores of molecular clouds and circumstellar envelopes of  late-type stars \citep{rei81}.  Masers have been used as  probes of gas with the H$_2$ number density of typically $10^4$--$10^{10}$ cm$^{-3}$.
For solar system objects, several maser and laser phenomena have been found; $e.g.$, ${\rm CO_{2}}$ (Venus and Mars: Mumma 1992) and ${\rm OH}$ (for many comets: {\it e.g.} Crovisier {\it et al.} 2002). Each phenomenon would be induced by  different physical processes.
While a thermal 22.235 GHz water line was possibly detected for comet Hale-Bopp (Bird {\it et al.} 1997), the first detection of ${\rm H_{2}O}$ maser in the solar system was reported at the catastrophic impact of comet Shoemaker-Levy9 and Jupiter (Cosmovici {\it et al.} 1996). This report suggests that such an incident can induce collisional pumping for water masers. Recently, Pogrebenko {\it et al.} (2009) (abbreviated as POG hereafter) have reported the detections of ${\rm H_{2}O}$ masers from the Saturnian satellites (Titan, Hyperion, Enceladus and Atlas) with the Medicina 32m and Mets${\rm \ddot{a}}$hovi 14m telescopes. This is interesting because, unlike a temporal phenomenon such as the break-up and disruption of a comet, we can perform long period monitoring of ${\rm H_{2}O}$ emission using ground-based telescopes, space telescopes ({\it e.g.} Herschel Space Telescope) and spacecrafts.

So far, we do not have much knowledge about the maser mechanism in the solar system, 
although a lot of water maser phenomena are observationally studied for extra-solar objects.
The combination of ground, space and in-situ observations would contribute to understand the nature of water maser emission if the presence in the Saturnian system is verified.

Therefore, we must accumulate more data of the water maser lines in the Saturnian system. 
In this letter, we report our trial of detecting water maser satellites with the 45m radio telescope at Nobeyama Radio Observatory (NRO). We observed the major Saturnian satellites for which POG reported the detections, and in addition, we observed a few other inner satellites.



\section{Observations}

We observed the Saturnian satellites with the NRO 45m telescope at the water maser frequency 22.23508 GHz on 15-17, 24, 27, 28 May in 2009. 
We used a cooled HEMT amplifier as the receiver front ends and an acousto-optical spectrometer (AOS-H) as back ends. The total band width of an acousto-optical spectrometer AOS-H was 40 MHz, and frequency resolution was 37 KHz, which corresponds to velocity resolution of $\sim 0.6$ km s$^{-1}$.
All the observations were made using the position switching method. We pointed the telescope toward the center of Titan and Saturn. 
A typical integration time of one scan was 20 sec. As the half-power beam width (HPBW) of the telescope ($\sim72^{"}$ at 23 GHz) was larger than the angular diameter of the Saturn ($41^{"}$ including the ring), we observed several satellites simultaneously. For example, we could observe Hyperion when we observed Titan on 16, 17, 18 and 28 May, and Enceladus and Atlas when we observed the Saturn throughout the observational days, though a slight offset of pointing resulted in the worse detection upper limit. Furthermore, we select Iapetus as the OFF points of the position switching. We segregated these satellites using the differences of Doppler shift frequency in the data analysis. Telescope pointing was checked with nearby strong H$_2$O maser stars. The antenna temperature, ${\rm T_{A}^{*}}$, was obtained by the chopper wheel method correction for atmospheric and antenna ohmic losses. A typical range of the system temperature was 120--300K, which depended on weather condition and airmass at the observations. Table 1 gives a summary of the observations.

We performed base line fitting with the 3rd order polynomial to eliminate the continuum emission from the sky and Saturn. Then, we co-added and binned these spectra with taking the Doppler shift for each object into consideration. Radial velocities of the Saturnian satellites were calculated using the JPL Horizons On-Line Ephemeris System (Giorgini {\it et al.} 1996). The resultant velocity resolution was about $\Delta v \simeq 1$  km  s$^{-1}$. These processes were done on the software package NEWSTAR, and procedures to obtain the final spectra after the Doppler correction were carried out on the software which had been developed for the solar system objects. 


\section{Results and Discussion}

Figure 1 shows the obtained spectra (upper two  and lower-left panels) and the illustration of the satellite positions (lower-left panel) at the time of observations for Titan, Hyperion and Iapetus. Figure 2 also shows the spectra (upper panels) and illustrations of positions (lower panels)  for Enceladus and Atlas. The position of a satellite $\theta$ is defined as an angle between the line of sight towards Saturn and the line of a satellite and Saturn ($i.e.$ $\theta=0$ when the object transits on the Saturn seen from the Earth). Table 2 shows a typical 1$\sigma$ upper limit of the signal considering the factor of a satellite position inside the telescope beam. 
The conversion efficiency of flux density to antenna temperature (${\rm T_{A}^{*}}$), 2.8 Jy ${\rm K^{-1}}$, was used for the calculations.
The observational results show the data combined on each day and the entire days during the observation period. 

POG reported that the water line detections of Titan and Hyperion were $\sim$30 mK ($3.8 \sigma$ for 8 hour integration) and $\sim$50 mK ($4.0 \sigma$ for 6 hours integration), respectively, which correspond to 300 mJy and 500 mJy, respectively. They found that these emissions were seen on almost the same satellite positions for both objects, and based on these results, they inferred a common mechanism of the emission such as Saturnian magnetosphere bow shock. On 16-18 and 28 May 2009, Titan and Hyperion were near conjunction so that we could observe them simultaneously in the telescope beam. During 16-18 May, especially, these objects were at almost the same position. Therefore, if the maser emissions had been caused by the common mechanism, we would have detected them simultaneously for these two objects. However, our observations this time could not detect the signals of the emissions, and all the data combined also showed no symptom.
We estimated typical values of the $3\sigma$ upper limit for each daily data. We also obtained $3\sigma$ upper limits during the observations using all the daily data combined. The typical(all) $3\sigma$ upper limits were 200(60) mJy for Titan and 240(80) mJy for Hyperion, which means that we could certainly detect the line emissions if the levels of flux densities were as high as those of POG.

Although POG did not report the water maser emission on Iapetus, we monitored this satellite during our observation period aiming this object at the OFF-point of the position switching observations. We could not find any signal stronger than $\sim$210(75) mJy ($3\sigma$ upper limit) .

The ${\rm H_{2}O}$ ice and vapor plume on Enceladus has been reported by the Cassini Ultraviolet Imaging Spectrometer (Hansen {\it et al.} 2006 and Hansen {\it et al.} 2008), and the hypothesis that there exists liquid water in the crust has been proposed (Porco {\it et al.} 2006). As for the water maser emission, the mass ratio of water vapor to ice in the plume is important because a majority of water molecules must be in the form of vapor in order to have maser emission originated from the Enceladus plume. A recent theoretical study indicates that the plume would be dominated by vapor from the thermodynamics perspective (Kieffer {\it et al.} 2009). The maser line intensity reported by POG was 500 mJy (4.2$\sigma$), and they estimated  the column density of water vapor from the observed maser intensity, which agrees with the column density of the water vapor plume observed in UV ($n{\rm =1.5 \times 10^{16}~cm^{-2}}$: Hansen {\it et al.} 2006). Nevertheless, we did not find  appreciable maser emission in our data. The $3\sigma$ upper limits were 780(540) mJy. Our data were worse compared with those of POG because Enceladus was located around the edge of the telescope beam much of the observational time. However, these results might indicate that the flux of the plume is varying, or the plume is sporadic; like a geyser.

POG reported the most certain detections for Atlas. The averaged spectrum showed a peak with 32 mK and S/N=7.0. From the satellite positions, they found that the maser emissions occurred on the trailing side which was several thousand km away from Atlas, and suggested that the disturbance of the Atlas's motion had caused the emission in the edge regions of Saturnian rings A and F.
We attempted to verify this subject in our data. We divided the positions of Atlas into 3 parts; $i.e.$, position 1 and 3: positions passing before the geometrical edges of the Saturn's rings seen from the Earth, and Position 2: those after the rings (see the positions of Atlas in Figure 2). For each position, we combined the data, and furthermore, we tested the data on 19 May because Atlas passed by the geometrical edges of the rings ($\theta=90^{\circ}$) during the observations. However, we could not have positive results for each case. All the combined data did not show any prominent feature of the emission. The 3$\sigma$ detection limits for each data set were in the range of 430--520(400) mJy.

We also checked several inner satellites which have a diameter larger than $\sim$10 km;  Mimas, Janus, Epimeteus, Prometheus, Pandora and Pan. For all the satellites, the acquired data did not show indicative of the maser emission. The typical value of $3\sigma$ upper limits were 500(210) mJy.

We had an opportunity to observe the Saturnian water maser line at the time of the ring's disappearance; that is, we see the Saturn ring almost in the edge-on view, and we obtained the data which had comparable or even better sensitivities than those by POG for almost of the satellites, though our observations were made for a limited period in 2009 May.  If maser emission is stationary in intensity at the levels as those reported by POG, we could detect them. However the results were negative for all of the observed Saturnian satellites. From these results, we conclude that the water maser  in the Saturnian system may be sporadic in nature and it is strongly restricted to the time and position of satellites. We have to monitor the satellites for  longer periods and to obtain statistical results. These studies would be useful to figure out the water maser emission reported in the Saturnian system.
In addition, we should also perform monitoring observations for other icy bodies; Jovian satellites, comets, outer asteroids and Kuiper belt objects would be inside the scope. 
As suggested by Cosmovici {\it et al.} (1996), the maser emission may be induced by catastrophic events. Such events like a disruption or eruption can occasionally be found among the solar system objects; ($e.g.$ C/1999 S4 Linear: disruption and dissipation, 29P/Schwassmann-Wachmann: outburst, 7968 Elst-Pizarro: impact or cometary activity (Toth 2000 and Hsieh et al. 2004), etc.). We should not miss events which will occasionally happen and carry out the observations to accumulate the data.




\section{Summary}

\begin{itemize}
\item A search for water maser emission (22.22351 GHz) from several Saturnian satellites; Titan, Hyperion, Enceladus  Atlas, Iapetus, and other inner satellites with a diameter larger than 10 km (Mimas, Janus, Epimeteus, Prometheus, Pandora and Pan) were carried out with the 45m radio telescope at Nobeyama radio observatory.
\item We could not confirm any emission line for all the satellites. The typical $3\sigma$ upper limits of daily(average) data for Titan, Hyperion, Enceladus and Atlas were 200(60), 240(80) 780(540) and 430-520(400) mJy, respectively, and for the other inner satellites, the S/N obtained were almost the same.
\item The sensitivities were comparable or even better than those of POG for most of the satellites.
\item From our observations, we infer that water maser emission in the Saturnian system is sporadic in nature.
\end{itemize} 



\begin{figure}
  \begin{center}
    \FigureFile(160mm,80mm){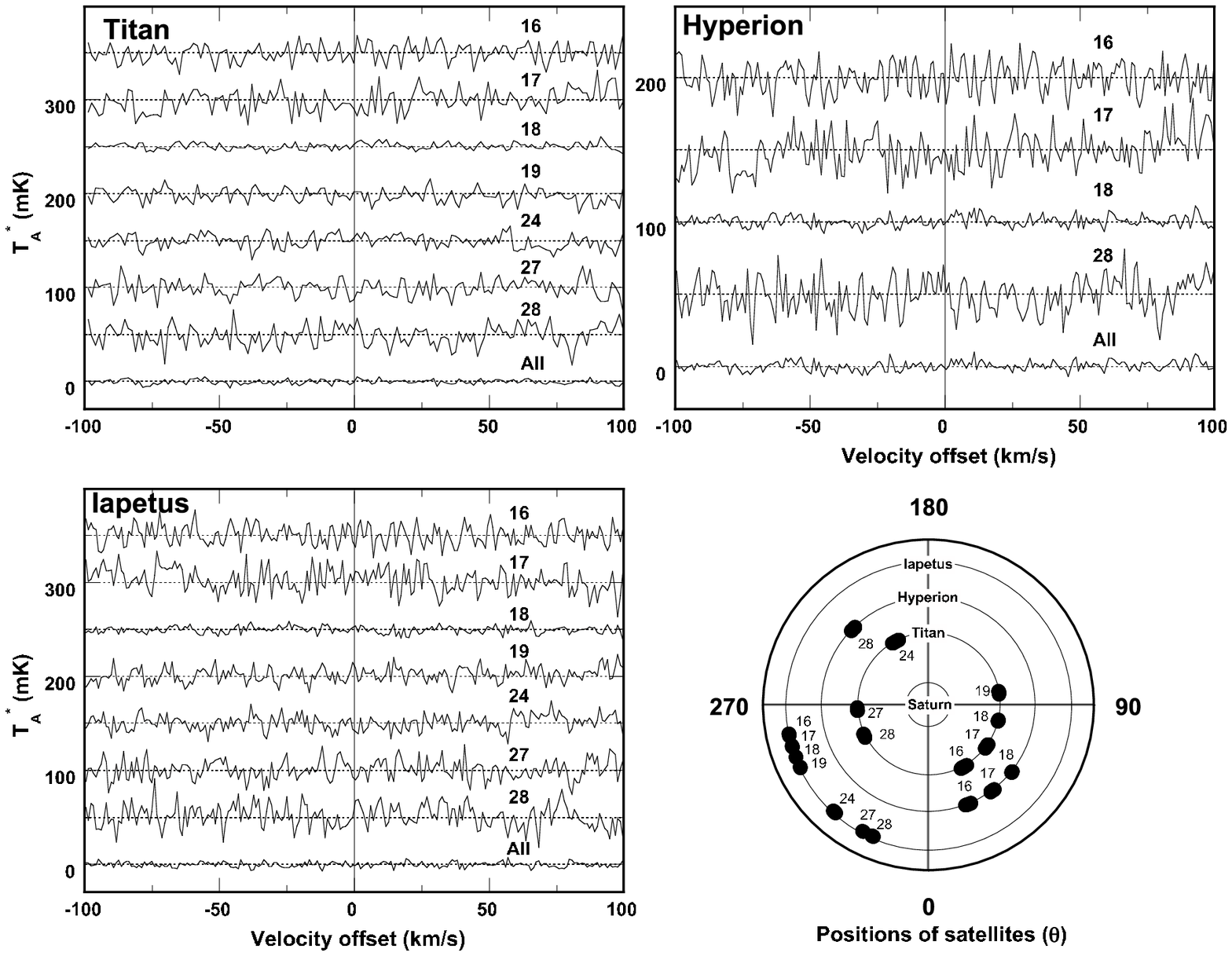}
  \end{center}
  \caption{Observational results and positions of satellite for Titan, Hyperion and Iapetus. Each data set has been shifted and plotted every 50 mK. Horizontal axis is target-centric velocity offset. The observed date in May 2009 (as shown in Table 1) is indicated  on each spectrum.
The position of a satellite $\theta$ is defined as an angle between the line of sight toward Saturn and the line of a satellite and Saturn ($i.e.$ $\theta=0$ when the object transits on the Saturn seen from the Earth). The radius in the position plot (lower-right panel) does not illustrate a real scale for each satellite.}\label{fig:sample}
\end{figure}

\begin{figure}
  \begin{center}
    \FigureFile(160mm,80mm){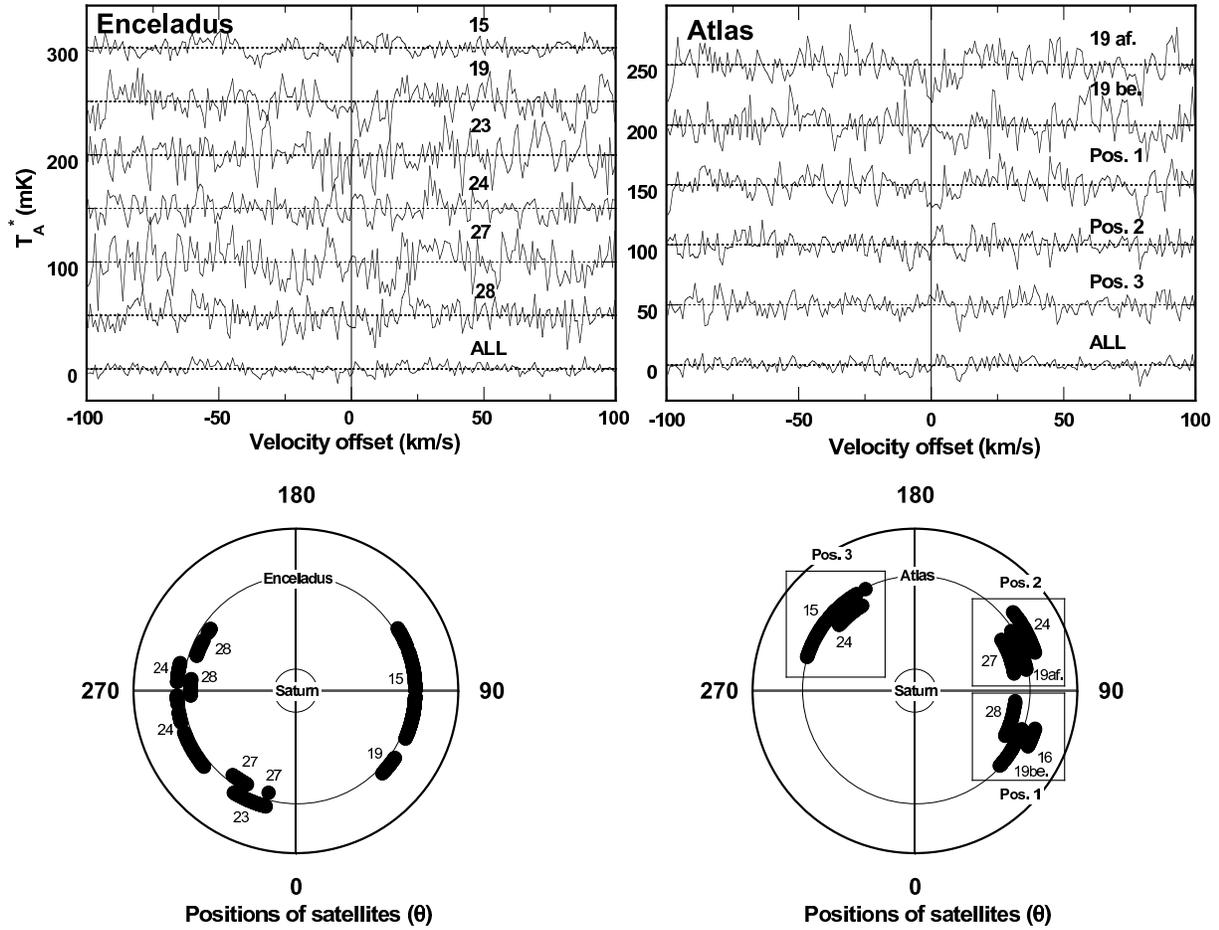}
  \end{center}
  \caption{Observational results and positions of satellites for Enceladus and Atlas. The labels 19af. and 19be. mean the combined results of 19 data ($\theta > 90^{\circ}$) and ($\theta < 90^{\circ}$), respectively.}\label{fig:sample}
\end{figure}


\begin{table}

\begin{minipage}{16cm}
  \caption{Observational Summary.}\label{tab1}
\begin{center}

    \begin{tabular}{rcccc}
      \hline
      Day    & Time (UT) &  IT (sec)\footnotemark[$*$]& IS (sec)\footnotemark[$\dagger$]\\
      \hline
      May 15 & 08:49-13:51 & 5260&-  \\
          16 & 05:49-11:12 & 3140 & - \\
          17 & 07:48-13:48 & 2180 & - \\
          18 & 09:28-10:44 & 2000 & 2000 \\
          19 & 08:44-12:37 & - & 1560 \\
          23 & 05:19-06:40 & 4000 & 3000 \\
          24 & 06:43-13:29 & 2000 & 2020 \\
          27 & 07:29-11:15 & 2000 & 2020 \\
          28 & 07:16-11:44 & 2000 & 2320 \\
\hline
       Total &  & 20580 & 17100 \\
      \hline
    \end{tabular}
\end{center}
\begin{center}
  \footnotetext[$*$]{ON-source integration time of Titan.}
  \footnotetext[$\dagger$]{ON-source integration time of Saturn.}
\end{center}
\end{minipage}

\end{table}

\begin{table}
\begin{minipage}{16cm}
  \caption{Observational Results.}\label{tab1}
  \begin{center}
    \begin{tabular}{cccc}
      \hline
      Satellite    & Date (or Data) & 1$\sigma$ U.L. (mJy)\footnotemark[$*$]\\
      \hline
      Titan & 16-19,24,27,28&65\\
            & All           &20     \\
      Hyperion & 16-18,28   &80 \\
            & All           &25     \\
      Iapetus & 16-19,24,27,28& 70\\
              & All           & 25\\
      Enceladus& 15,19,23,24,27,28&260\\
             & All &  180 \\
      Atlas  &Pos.1,2,3&170,140,170\\
             &19be.,19af.&140,150\\
             &All        & 130\\
      Other satellites &16-19,24,27,28&170\\
             &             &70\\
      \hline
    \end{tabular}
  \end{center}

\footnotetext[$*$]{A typical 1$\sigma$ upper limit of the signal considering the factor of a satellite position inside the telescope beam. The observational results show the data combined on each day and the entire days during the observation period. For Atlas, we show the results at each position.}
\end{minipage}
\end{table}


\bigskip

We are grateful to Dr. J. Crovisier, the referee of this paper for suggestions to improve the manuscript, and S. Takano and J. Maekawa for advice and discussions.
This research made use of ephemeris given by NASA Jet Propulsion Laboratory, California Institute of Technology.

%
%


\end{document}